\documentclass[preprint,prb,aps,draft,a4paper] {revtex4}
\usepackage[final]{graphics}
\usepackage{epsf}
\def\CASPT2{ANDERSSON:90,ANDERSSON:92}
\def\MSCASPT2{FINLEY:98,ZAITSEVSKII:95}

\def\abinitio{{\it ab initio}}

\def\cmm1{~cm$^{-1}$}
\def\etal{{\em et al.}}

\def\nua1g{$\bar{\nu}_{a_{1g}}$}
\def\nue{$\bar{\nu}_{a_{1g}}$}

\def\Re{$R_e$}

\def\DRet2g{$\Delta R_e(t_{2g}-f)$}

\def\Cethp{Ce$^{3+}$}

\def\Vthp{V$^{3+}$}
\def\Ythp{Y$^{3+}$}

\def\CstNaYCl6{Cs$_2$NaYCl$_6$}
\def\CstNaLnCl6{Cs$_2$NaLnCl$_6$}
\def\CstNaYBr6{Cs$_2$NaYBr$_6$}
\def\KtNaGaF6{K$_2$NaGaF$_6$}
\def\KtNaScF6{K$_2$NaScF$_6$}
 
\def\CsCaF3{CsCaF$_3$}
\def\KCdF3{KCdF$_3$}
\def\KMgF3{KMgF$_3$}
\def\KZnF3{KZnF$_3$}

\def\KtCrO4{K$_2$CrO$_4$}
\def\KtFeO4{K$_2$FeO$_4$}
\def\KtSO4{K$_2$SO$_4$}
\def\KtSeO4{K$_2$SeO$_4$}

\def\CstGeF6{Cs$_2$GeF$_6$}
\def\CstZrCl6{Cs$_2$ZrCl$_6$}
\def\CstUCl6{Cs$_2$UCl$_6$}
\def\CstUBr6{Cs$_2$UBr$_6$}

\def\TMAtUCl6{(TMA)$_2$UCl$_6$}
\def\CstUOtCl4{Cs$_2$UO$_2$Cl$_4$}

\def\CeClsthm{(CeCl$_6$)$^{3-}$}

\def\fd{$f^n\ \rightarrow\ f^{n-1}d^1$}

\begin{document}

   \title{ 
           Prediction of pressure-induced red shift of 
           $f^1 \rightarrow\ d(t_{2g})^1$
           excitations in Cs$_2$NaYCl$_6$:Ce$^{3+}$
           and its connection with 
           bond length shortening.
         }
   \author{
           Fernando Ruip\'erez,$^a$
           Luis Seijo,$^{b}$
           and 
           Zoila Barandiar\'an$^{b}$\thanks{ Corresponding author.
                       Electronic mail: zoila.barandiaran@uam.es }
          }
   \affiliation{
           $^{a,b}$Departamento de Qu\'{\i}mica, C-XIV,
           Universidad Aut\'onoma de Madrid, 28049 Madrid, Spain.\\
           $^b$Instituto Universitario de Ciencia de Materiales Nicol\'as Cabrera,
           Universidad Aut\'onoma de Madrid, 28049 Madrid, Spain
           }
   \date{\today}
   \begin{abstract}
Quantum chemical calculations including embedding,
scalar relativistic, and dynamic electron correlation
effects on Cs$_2$NaYCl$_6$:(CeCl$_6$)$^{3-}$ embedded
clusters predict ($i$) red shifts of the
\mbox{$f^1 \rightarrow\ d(t_{2g})^1$} transition with pressure and
($ii$) bond length shortening upon $f \rightarrow\ d(t_{2g})$
excitation. 
Both effects are found to be connected which suggests that
new high pressure spectroscopic experiments could
reveal the sign of the bond length change.
   \end{abstract}
   \maketitle
\newpage
\section{INTRODUCTION}
\label{SEC:Introduction}

Recent {\it ab initio} theoretical studies of the structure and spectroscopy 
of octahedral complexes of lanthanide/actinide ions
have shown that the bond length between the \mbox{$f$-element} 
and the surrounding ligands shortens upon 
$f^n \rightarrow \ f^{n-1}d(t_{2g})^1$ 
excitation.~\cite{SEIJO:01:b,SEIJO:03:c,BARANDIARAN:03:a,BARANDIARAN:03:b,BARANDIARAN:SUB}
This is so for different
halide ligands (F, Cl, Br), environments
(crystals, liquid solutions, gas phase), oxidation states
of the $f^n$ ion (III, IV), and number of $f$ electrons
(across the lanthanide/actinide 
series).~\cite{SEIJO:01:b,SEIJO:03:c,BARANDIARAN:03:a,BARANDIARAN:03:b,BARANDIARAN:SUB,RUIPEREZ:IP}
Yet, it contradicts the widespread assumption
that \fd\ excitations lengthen the impurity-ligand
bonds.~\cite{ANDRES:96,SHEN:98,KHAIDUKOV:00,DEREN:00}

Usually, the materials involving $f$-element ion complexes are        
liquid or  solid solutions and                                       
a direct experimental proof of the bond shortening
could be obtained, in principle, by means of ground ($f^n$) and
excited state ($f^{n-1}d^1$)
EXAFS measurements. 
In this respect,                                                    
a theoretical study of 
(CeX$_6$)$^{3-}$ (X= F, Cl, Br)
complexes in cubic elpasolites,
liquid acetonitrile solution, and gas 
phase,
has pointed out the chloride and bromide complexes
in acetonitrile as good candidates for 
excited state EXAFS experiments because the negative
bond length shifts are largest and the liquid medium
could favor the experimental setting.~\cite{BARANDIARAN:SUB}
However,
excited state EXAFS measurements are extremely
demanding because the difficulties inherent to
EXAFS experiments and
their interpretation are
extended by the need of pumping the samples
onto the excited $f^{n-1}d^1$ levels for long periods of time
and by 
the uncertainties on the actual excited
state population attained. 
As a matter of fact, no successful experiments of this
kind have yet been reported  in octahedral complexes
of $f$ element ions in solid or liquid media,
to our knowledge.

In this paper, on the basis of quantum chemical simulations,
we propose that spectroscopic studies under high hydrostatic 
pressure~\cite{DRICKAMER:65,BRAY:01}
are an alternative, relatively simple
experimental technique to proof or reject the predicted
bond length shortening upon 
$f^n \rightarrow \ f^{n-1}d(t_{2g})^1$ excitation.
In effect,  quantum chemical calculations 
in \CstNaYCl6:\Cethp\ reveal that 
the shortening in bond length 
upon $f^n \rightarrow \ f^{n-1}d(t_{2g})^1$ excitation
results in a continuous redshift of the transition energy
with pressure in the range 1 bar -- 26 kbar. 
More generally,
the sign of the bond length
change upon excitation determines the sign of
the slope of the transition energy with pressure.
As a consequence,
the theoretically found sequence of bond lengths 
in octahedral complexes,~\cite{BARANDIARAN:03:b}
\Re[$f^{n-1}d(t_{2g})^1$] $\le$
\Re[$f^n$] $\le$
\Re[$f^{n-1}d(e_g)^1$],
consistently leads to
deacreasing 
\mbox{$f^n$ $\rightarrow$ $f^{n-1}d(t_{2g})^1$},
constant  \mbox{$f^n$ $\rightarrow$ $f^n$}, and
increasing \mbox{$f^n$ $\rightarrow$ $f^{n-1}d(e_{g})^1$}
transition energies 
with hydrostatic pressure.
%
These results also enhance the value of high pressure experiments
to alter the relative positions of excited 
states~\cite{SHEN:98,DRICKAMER:65,YOO:91} and to help
the usually difficult assignment of the 
$f^{n-1}d(e_{g})^1$ highest energy
levels.~\cite{TANNER:03}


\section{RESULTS AND DISCUSSION}
\label{SEC:Results}

The effects of hydrostatic pressure on the local structure
and electronic transitions of 
the \CeClsthm\ defects in \CstNaYCl6\ 
were studied using the relativistic
\abinitio\ model potential (AIMP)
embedded cluster method.~\cite{BARANDIARAN:88,SEIJO:99}
More detailed descriptions of the method, as applied
to $f$ element ions in ionic crystals, can be found in 
Refs.~\onlinecite{SEIJO:03:c} and~\onlinecite{BARANDIARAN:03:b}.
The effective core potentials used for 
Ce~~\cite{SEIJO:03:a} 
and Cl~~\cite{BARANDIARAN:92}
in an experimental and theoretical
study of the absorption and emission spectra of
\Cethp\ in elpasolite lattices~\cite{TANNER:03}
were also used here to incorporate scalar relativistic
effects. 
In order to include dynamic electron correlation, SCF
calculations were performed on the 4$f^1$ and 5$d^1$ electronic
states and were used as
reference for second order perturbation
calculations  where 57 valence electrons were
correlated that occupy molecular orbitals
with main character Ce 5$s$, 5$p$, 4$f$/5$d$ and
Cl 3$s$, 3$p$.~\cite{ANDERSSON:90,ANDERSSON:92}
(The SCF results of the structure  and pressure-induced shifts
of electronic transitions 
are in qualitative agreement with those
presented here; therefore, the predictions made
in this paper already hold qualitatively at that level of
theory.)
The effects of pressure were modelled by using
the AIMP embedding potentials produced in a recent
study of high pressure effects on the
structure and spectroscopy of \Vthp\ defects
in \CstNaYCl6.~\cite{SEIJO:03:b} 
The embedding potentials corresponding to a
given lattice constant were obtained through
self-consistent embedded ions calculations;~\cite{SEIJO:03:b}
they allow to incorporate quantum mechanical 
interactions with the host \CstNaYCl6\ ions
in the \CeClsthm\ cluster 
Hamiltonian.~\cite{BARANDIARAN:88,SEIJO:99}
The values of the lattice constant used are listed
in Table~\ref{TAB:ONE}, where
the corresponding $-\Delta V/V$ values are also
shown. These data define a
pressure range \mbox{1~bar~--~26~kbar} 
(see Table~\ref{TAB:ONE}) if the isothermal compressibility of the
material is \mbox{$\kappa = (-\Delta V/V)/\Delta P = 
4\times$10$^{-3}$~kbar$^{-1}$.~\cite{SEIJO:03:b}}

Since \Cethp\ substitute for \Ythp\ ions that occupy $O_h$
sites 
the embedded cluster energies
of all electronic states in the 
$f^1$ (the ground state $^2A_{2u}$, $^2T_{2u}$, and $^2T_{1u}$) and
$d^1$ ($^2T_{2g}$ and $^2E_g$) configurations were calculated
at different Ce-Cl distances, R,
for each lattice compression
(there has been no reports of significant coupling
to Jahn-Teller active normal modes).
The plots of the potential energy surfaces for ambient pressure
and 26~kbar are given in Fig.~\ref{FIG:PES}.
We obtained the
equilibrium distances, \Re, and totally symmetric
vibrational frequencies, \nue, as
in Ref.~\onlinecite{SEIJO:03:b}; the vertical ($\Delta E$)
and minimum-to-minimum ($\Delta E_e$) transition energies,
and the shifts of bond length that occur upon
excitation ($\Delta R_e$) were also obtained and are presented
in Table~\ref{TAB:ONE} and Figures~\ref{FIG:PES}
to \ref{FIG:DeltaEvsR}.

The results in
Table~\ref{TAB:ONE} and Fig.~\ref{FIG:PES} show
that the  Ce$-$Cl bond length shortens upon
$f^1~^2A_{2u} \rightarrow$~d$^1~^2T_{2g}$
excitation at ambient pressure and at higher
pressures as well. Figures~\ref{FIG:PES} and
\ref{FIG:Peffects}  reveal a
redshift of the 
$f^1~^2A_{2u} \rightarrow$~d$^1~^2T_{2g}$
transition with pressure.
Both observations are connected:
the sign of the bond length shift determines
the sign of the slope of the electronic transition
with pressure, being  both of them negative, for
the $f^1~^2A_{2u} \rightarrow$~d$^1~^2T_{2g}$ 
transition. This connection can be established
as follows. 
On the one hand (see Fig~\ref{FIG:DeltaEvsR}),
the slope of the 
calculated transition energy $\Delta E$
with the Ce$-$Cl distance R, $d\Delta E/dR$,
varies little with pressure (see solid lines) and has the same
sign as (and is very similar to) the slope of the
minimum-to-minimum transition energy
$\Delta E_e$ with
the Ce$-$Cl equilibrium distance of the
ground state,
$d\Delta E_e/dR_e(^2A_{2u})$. The same is true
for the slope of the vertical, Frank-Condon transition
$\Delta E[R_e(^2A_{2u})]$ with
$R_e(^2A_{2u})$,
$d\Delta E[R_e(^2A_{2u})]/dR_e(^2A_{2u})$. Furthermore, 
the observations and conclusions drawn from 
Fig~\ref{FIG:DeltaEvsR} also apply
to any other $i \rightarrow f$ transition 
studied here, so that 
$$
   {\rm sign\  of\ \ }\frac{d\Delta E(i \rightarrow f)}{dR}\ \ = \ \ 
   {\rm sign\  of\ \ }\frac{d\Delta E_e(i \rightarrow f)}{dR_{e,i}}\ ,
   \eqno{(1)}
$$
and their magnitude is similar.
On the other hand, it has been shown~\cite{BARANDIARAN:SUB}
that when the force constant of the initial 
and final states are similar, 
$k_i \approx k_f \approx k$, like it is the case here
(note that \nue\ values are given in Table~\ref{TAB:ONE}
and
$k = M_{Cl}4\pi^2$\nue$^2c^2$
for the ${a_{1g}}$ normal mode)
the slope of the transition
energy $d\Delta E(i \rightarrow f)/dR$
depends basically on the force constant
$k$
and on the bond length shift 
$\Delta R_e(i \rightarrow f) =
R_e(f) - R_e(i)$,
accompanying
the transition, as follows:~\cite{BARANDIARAN:SUB}
\mbox{$
   d\Delta E(i \rightarrow f)/dR \approx
    - k \, \Delta R_e(i \rightarrow f)\,.
$}
This,  together with the previous observations (Eq. 1),
and the fact that $dR_{e,i}/dP \le 0$,
leads to the conclusion that the sign of the
slope of a transition with pressure coincides
with the sign of the bond length shift that
occurs upon excitation:
$$
   {\rm sign\  of\ \ }\frac{d\Delta E_e(i \rightarrow f)}{dP} =\ 
   {\rm sign\  of\ \ }\Delta R_e(i \rightarrow f)\,,
   \eqno{(2)}
$$
and the same is true for the Frank-Condon transition.
In effect (Fig~\ref{FIG:Peffects}, Table~\ref{TAB:ONE}), 
the \mbox{$f\rightarrow f$} transitions (with $\Delta R_e \approx 0$)
are quite insensitive to pressure,
$f\rightarrow d(t_{2g})$ transitions (with $\Delta R_e < 0$)
shift to the red, and
$f\rightarrow d(e_g)$ excitations (with $\Delta R_e > 0$)
shift to higher energies with pressure.

It is known, from transition metal spectroscopy,
that the so called
10$D_q$ transition,~\cite{SUGANO:70}
$d(t_{2g})$$\rightarrow$$d(e_g)$,
shifts to high energy values with 
pressure,~\cite{DOLAN:86,SEIJO:93}
which is related to the bond length increase
upon $d(t_{2g})$$\rightarrow$$d(e_g)$
excitation (Eq. 2); the same behavior is obtained here for 
$f^{n-1}d(t_{2g})^1$ $\rightarrow$ $f^{n-1}d(e_g)^1$
(Table~\ref{TAB:ONE}, Fig~\ref{FIG:Peffects}). It should be noticed,
however, that this information
alone does not suffice to infer the effects of pressure
on the $f\rightarrow d$ transitions; rather,
the position of both potential
energy surfaces $f^{n-1}d(t_{2g})^1$ and 
$f^{n-1}d(e_g)^1$ 
relative to the potential energy surface
of the $f^n$ ground state in the R axis of the configurational diagram,
has to be also known.~\cite{BARANDIARAN:03:b}
These relative positions have been found to be
\Re[$f^{n-1}d(t_{2g})^1$] $\le$
\Re[$f^n$] $\le$
\Re[$f^{n-1}d(e_g)^1$] by calculations here and
elsewhere, as commented
above,~\cite{BARANDIARAN:03:b}
and have been interpreted as resulting from 
the following simple model of interactions:
The inner lanthanide (Ln) (or actinide, An) $f^n$ open-shell electrons are
shielded from the ligands by the outer Ln 5$p^6$
(An 6$p^6$) closed-shell electrons, whose interactions
with the ligands determine the bond distance in states
of $f^n$ configuration. Instead, upon $f^n$$\rightarrow$$f^{n-1}d^1$
excitation one electron has crossed the 
5$p^6$ (6$p^6$) barrier 
and becomes exposed to covalent interactions with
the ligands, at the same time that it leaves a 4$f$ (5$f$) hole 
behind,
available for charge transfer from the ligands, both 
of which contribute to shortening the bond length
between the $f^{n-1}d^1$ baricenter and the ligands.
Finally, a large $d$($t_{2g}$)-$d$($e_g$) octahedral
ligand splitting
shortens the $f^{n-1}d(t_{2g})^1$ bond lengths
clearly below the $f^n$ ones ($\Delta $\Re~$<$0) and raises the
$f^{n-1}d(e_g)^1$ ones clearly above them ($\Delta $\Re~$>$0).~\cite{BARANDIARAN:03:b}
These bond length shifts ultimately lead to
the predicted red shift of $f^1$$\rightarrow$$d(t_{2g})^1$
and increase of 
$f^1$$\rightarrow$$d(e_g)^1$ transitions with pressure
in \CstNaYCl6:\Cethp.
Should the sequence of bond lengths 
be that traditionally assumed:
\mbox{\Re[$f^n$] $\le$ \Re[$f^{n-1}d(t_{2g})^1$] $\le$ \Re[$f^{n-1}d(e_g)^1$],}
both transitions
would increase with pressure.
Whether the former or the latter sequence of
bondlengths actually occurs could be revealed
by 
performing high pressure spectroscopic
experiments on this model system \CstNaYCl6:\Cethp\
in the range 1 bar -- 26 kbar.

From all the results we have discussed here it is possible to derive a
rule of thumb that can be applied to 
predict the sign of the change of a transition with
pressure. It consists of using the calculated configurational
diagram for ambient pressure 
(dashed lines in Fig~\ref{FIG:PES},
for instance)
and use it assuming that the effects of pressure
simply correspond to moving inwards across the R axis.
This approximation applies as a consequence of 
Eq. (1), which is a manifestation that the major
effect of pressure on these local electronic transitions
is a moderate decrease 
of the f-element -- ligand distance.

\section{CONCLUSIONS}
\label{SEC:Conclusions}

Previous \abinitio\ wavefunction based studies
of the structure and spectroscopy of octahedral
complexes of $f$ element ions in crystals, liquid
solutions, and gas phase have shown       that
the bond length between the $f$-element and the
ligands shortens upon 
$f^n \rightarrow \ f^{n-1}d(t_{2g})^1$ excitation.
This result contradicted the widespread assumption
that the bond length is longer in the  $f^{n-1}d^1$ states.
No experimental proof of the actual sign of the
bond length change upon excitation has been given.
We have presented here
a quantum chemical simulation
of spectroscopic experiments under high
hydrostatic pressure. The simulation includes
scalar relativistic effects, 
second order perturbation treatment of the valence electron
correlation, and quantum mechanical embedding
techniques. A continuous redshift of the lowest 
$f^n \rightarrow \ f^{n-1}d^1$ band
of \CstNaYCl6:\Cethp\ with pressure is predicted
in the range 1~bar to 26~kbar.
The  pressure-induced redshift is shown to be
related to the bond length shortening
that accompanies the $f\rightarrow d(t_{2g})$ excitation. 
Therefore, the simulation points to the
experimental detection of the pressure induced redshift 
as a way to support (or reject) the bond length shortening.
More generally, the calculations suggest that the sign
of the bond length change upon excitation determines
the sign of the pressure-induced shift of the transition, and the effects of
pressure on the $f \rightarrow f$ (no shift) and $f \rightarrow d(e_g)$
(shift to higher energies) transitions are also predicted
and are related to the respective negligible and
positive changes in bond length.
The \CstNaYCl6:\Cethp\ crystal is pointed out as an
ideal model system for the counterpart high pressure
experimental studies because the
higher $f^1\rightarrow d(e_g)^1$ band has been observed
below the cutoff of the host absorption.
\section*{ACKNOWLEDGEMENTS}
    This research was supported in part by 
    Ministerio de Educaci\'on y Ciencia, Spain,
    under contract BQU2002-01316,
    and Accciones de Movilidad PR2003-0024 and PR2003-0027.
    We are very grateful to Professor Norman M. Edelstein
    for stimulating discussions.
\bibliographystyle{prsty}

\widetext
\begin{table}[p]
\caption[]{
\label{TAB:ONE}
         Effects of high pressure on the potential energy surfaces of 
         4$f^1$ and 5$d^1$ manifolds of \CstNaYCl6:\Cethp.
         Distances in \AA, vibrational frequencies and energies in \cmm1.
}
\begin{ruledtabular}
\begin{tabular}{lrrrrrr}
\noalign{\smallskip}
\multicolumn{2}{r}{lattice constant, $a$} & 10.7396 & 10.6752 & 10.5463 & 
                   10.4067 & 10.3530 \\
 & $-\Delta V/V$ & 0 &  0.0179 &  0.0530 &  0.0902 &  0.1042 \\
 & pressure \footnote{Assuming $(-\Delta V/V)/P = 4 \times 10^{-3}$~kbar$^{-1}$. }
 & 1~bar & 4.5~kbar & 13.2~kbar & 22.5~kbar & 26~kbar \\
\cline{3-7}
\noalign{\smallskip}
 $R_{e}$         
     & 4$f^1$ $ ^2A_{2u}$      & 2.687  & 2.680  & 2.664  & 2.646  & 2.638  \\
     &        $ ^2T_{2u}$      & 2.688  & 2.681  & 2.665  & 2.647  & 2.639  \\
     &        $ ^2T_{1u}$      & 2.690  & 2.683  & 2.667  & 2.649  & 2.641  \\
     & 5$d^1$ $ ^2T_{2g}$      & 2.645  & 2.639  & 2.625  & 2.609  & 2.602  \\
     &        $ ^2E_{g }$      & 2.705  & 2.698  & 2.682  & 2.662  & 2.655  \\
     & $\langle{\rm 5}d^1\rangle$ \footnote{$R_e$ of 5$d^1$ center of gravity = 
                 $\left[3\times R_e(^2T_{2g}) + 2\times R_e(^2E_g)\right]/5$}
                               & 2.669  & 2.663  & 2.648  & 2.630  & 2.526  \\
\noalign{\smallskip}
 $\Delta R_{e}$  
     & 4$f^1$ $^2A_{2u}$$ \rightarrow\ $4$f^1$ $^2T_{2u}$  
                               &  0.001 &  0.001 &  0.001 &  0.001 &  0.001 \\
     &             $^2T_{1u}$  &  0.003 &  0.003 &  0.003 &  0.003 &  0.003 \\
     &                  $ \rightarrow\ $5$d^1$ $^2T_{2g}$  
                          & --0.042 & --0.041 & --0.039 & --0.037 & --0.036 \\
     &             $^2E_{g }$  &  0.018 &  0.018 &  0.018 &  0.016 &  0.017 \\
     &                  $ \rightarrow \langle{\rm 5}d^1\rangle$
                          & --0.018 & --0.017 & --0.016 & --0.016 & --0.015 \\
     & 5$d^1$ $ ^2T_{2g}$$\rightarrow$ 5$d^1$ $^2E_{g }$  
                               &  0.060 &  0.059 &  0.057 &  0.053 &  0.053 \\
\noalign{\smallskip}
$\overline{\nu}_{a_{1g}}$ 
     & 4$f^1$ $ ^2A_{2u}$      & 306    & 311    & 322    & 335    & 341    \\
     & $ ^2T_{2u}$             & 306    & 311    & 323    & 336    & 342    \\
     & $ ^2T_{1u}$             & 307    & 312    & 323    & 336    & 342    \\
     & 5$d^1$ $ ^2T_{2g}$      & 307    & 312    & 323    & 336    & 341    \\
     &        $ ^2E_{g }$      & 300    & 305    & 317    & 330    & 336    \\
\noalign{\smallskip}
 $\Delta E_e$                     
     & 4$f^1$ $^2A_{2u}$$ \rightarrow\ $4$f^1$ $^2T_{2u}$
                               &    460 &    460 &    480 &    490 &    500 \\
     &             $^2T_{1u}$  &    890 &    910 &    950 &   1000 &   1020 \\
     &                  $ \rightarrow\ $5$d^1$ $^2T_{2g}$
                               &  24300 &  24100 &  23700 &  23200 &  23000 \\
     &             $^2E_{g }$  &  47200 &  47300 &  47700 &  48100 &  48200 \\  
     &               $ \rightarrow \langle{\rm 5}d^1\rangle$ \footnote{$\Delta 
E_e$ of 4$f^1$~$^2A_{2u}$ $\rightarrow$ 5$d^1$ center 
           of gravity = $\left[3\times \Delta E_e(^2A_{2u}\rightarrow ^2T_{2g}) 
                 + 2\times \Delta E_e(^2A_{2u}\rightarrow ^2E_g)\right]/5$}
                               &  33460 &  33380 &  33300 &  33160 &  33080 \\
     & 5$d^1$ $ ^2T_{2g}$$\rightarrow$ 5$d^1$ $^2E_{g }$
                               &  22900 &  23200 &  24000 &  24900 &  25200 \\  
\end{tabular}
\end{ruledtabular}
\end{table}


     \widetext                                                
\newpage
\begin{figure}
  \caption{
  \label{FIG:PES}
           Effects of pressure on the potential energy surfaces
           of 4$f^1$ and 5$d^1$ electronic states of
           \CstNaYCl6:\Cethp. High pressure curves (solid lines)
           correspond to $-\Delta V/V =$ 0.1042 (26 kbar, if
           \mbox{$\kappa = (-\Delta V/V)/P$ = 4$\times$10$^{-3}$ kbar$^{-1}$).}
          }
\end{figure}

\begin{figure}
  \caption{
  \label{FIG:Peffects}
           Effects of pressure on the  4$f^1$$\rightarrow$4$f^1$
           and 4$f^1$$\rightarrow$5$d^1$ electronic transitions
           of \CstNaYCl6:\Cethp.
          }
\end{figure}

\begin{figure}
  \caption{
  \label{FIG:DeltaEvsR}
           Solid lines:
           4$f^1$ $^2A_{2u}$$ \rightarrow$ 5$d^1$ $^2T_{2g}$
           transition energies versus Ce-Cl distance at different
           pressures (1~bar and 4.5, 13.2, 22.5, and 26 kbar, respectively, 
           starting from the upper line). 
           Frank-Condon transition energies,
           E[$^2T_{2g}$;\Re($^2A_{2u}$)]
           $-$ E[$^2A_{2u}$;\Re($^2A_{2u}$)] (circles) 
           and
           minimum to minimum energies,
           E[$^2T_{2g}$;\Re($^2T_{2g}$)]
           $-$ E[$^2A_{2u}$;\Re($^2A_{2u}$)] (squares), 
           versus
           \Re($^2A_{2u}$) values at different 
           pressures.
          }
\end{figure}
\vfill


\begin{flushleft}
\thispagestyle{empty}
            \epsfysize = 20cm \epsfxsize = 15cm
            \epsffile{ 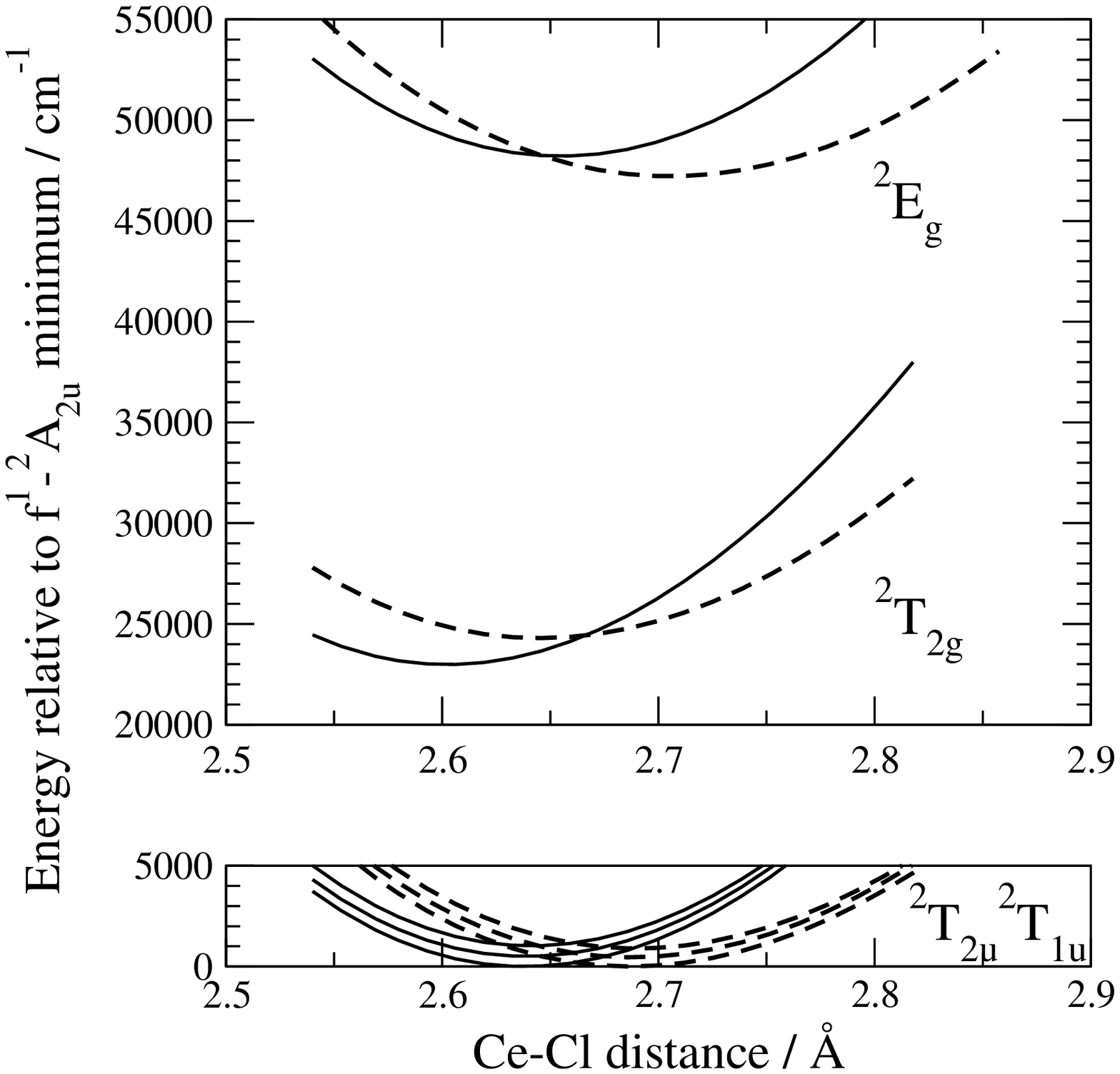 }

\vfill

Figure 1. Ruip\'erez \etal\  Journal of Chemical Physics
\newpage
            \epsfysize = 20cm \epsfxsize = 15cm
            \epsffile{ 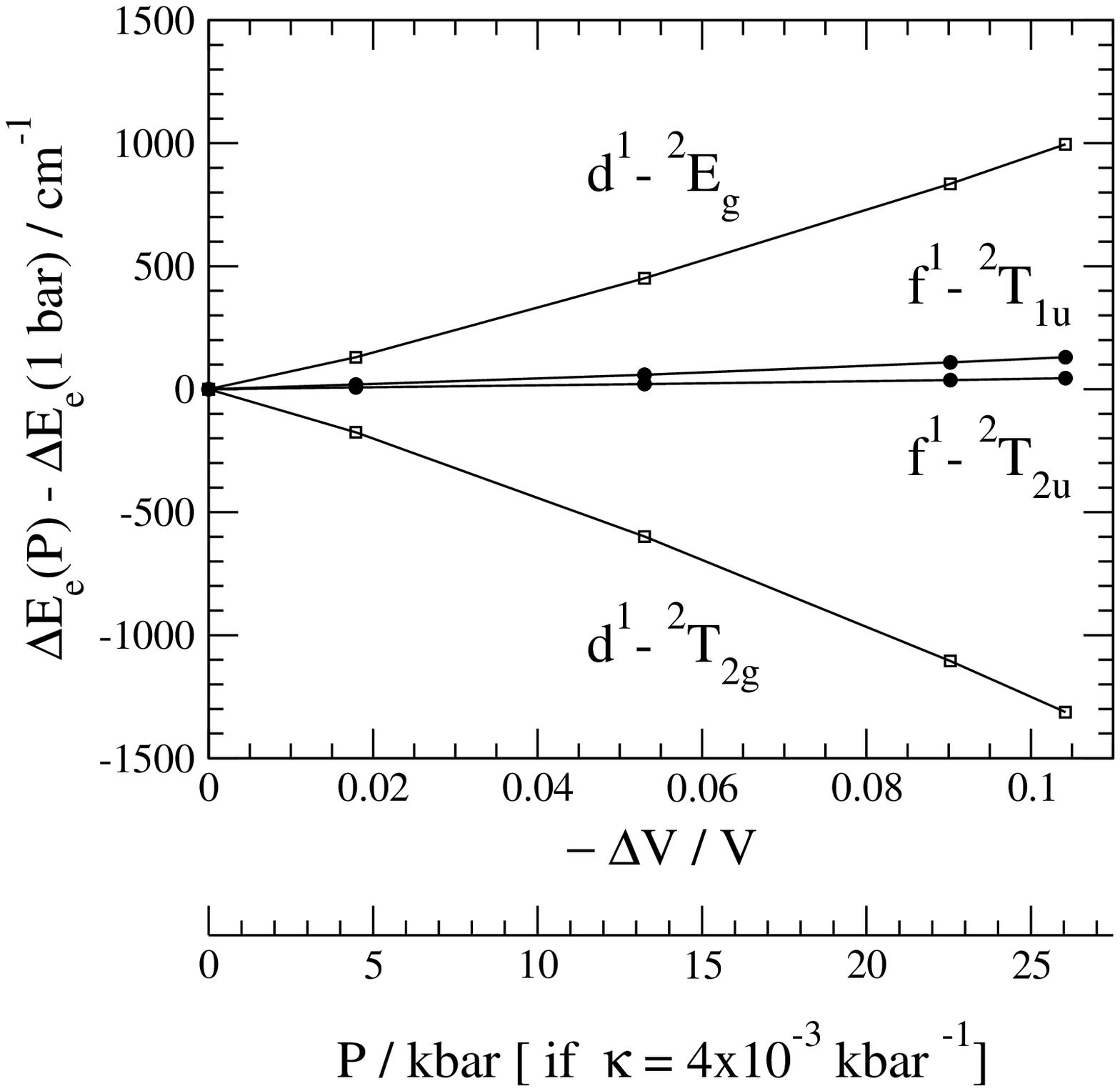 }
\vfill

Figure 2.  Ruip\'erez, \etal\  Journal of Chemical Physics
\newpage
            \epsfysize = 21cm \epsfxsize = 15cm
            \epsffile{ 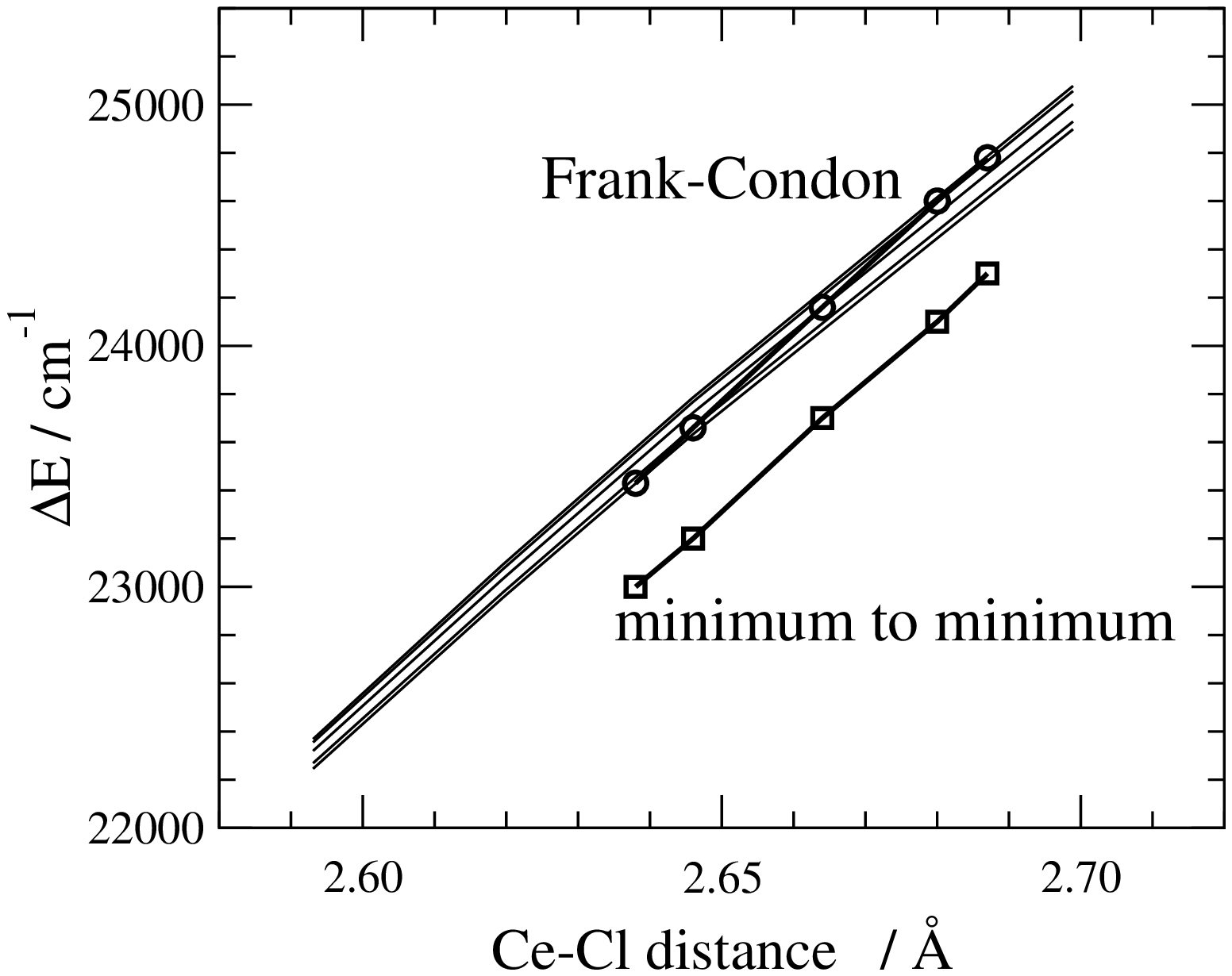 }
\vfill
 
Figure 3.  Ruip\'erez, \etal\ Journal of Chemical Physics
\end{flushleft}

\end{document}